\documentclass[12pt]{article}
\usepackage{graphicx}
\usepackage{cite}
\usepackage[centertags]{amsmath}
\usepackage{amsfonts}
\usepackage{amssymb}
\usepackage{amsthm}
\usepackage{pstricks}
\usepackage{color}
\usepackage{slashed}
\usepackage{newlfont}
\usepackage{ulem}
\usepackage{amsmath,amssymb}
\usepackage{array}
\usepackage{slashed}

\usepackage{float}
\usepackage{amsmath,amsfonts,graphicx,bm}

\usepackage{amsmath,amsfonts,graphicx,bbm,multicol}

\usepackage{subfigure}

\newcommand\rep\mathbf

\def\beq{\begin{equation}}
\def\eeq{\end{equation}}
\def\bea{\begin{eqnarray}}
\def\eea{\end{eqnarray}}

\def\bt{\begin{table}}
\def\et{\end{table}}
\def\bc{\begin{center}}
\def\ec{\end{center}}
\def\bi{\begin{itemize}}
\def\ei{\end{itemize}}
\def\bea{\begin{eqnarray}}
\def\eea{\end{eqnarray}}
\def\beas{\begin{eqnarray*}}
\def\eeas{\end{eqnarray*}}

\def\nn{\nonumber}

\begin{document}
\setcounter{page}{0}
\thispagestyle{empty}
\vspace*{-1.0cm}
\begin{flushright}
OSU-HEP-12-01
\end{flushright}
\vspace{0.1cm}

\begin{center}
{\Large\sc Diquark resonance and single top production at the Large Hadron Collider }

\vspace*{1.0cm} Durmus  Karabacak\footnote{Email address: durmas@ostatemail.okstate.edu},
                S. Nandi\footnote{Email address: s.nandi@okstate.edu}
        and     Santosh Kumar Rai\footnote{Email address: santosh.rai@okstate.edu}

\vspace*{0.2cm}
{\it Department of Physics and Oklahoma Center for High
Energy Physics, \\ Oklahoma State University, Stillwater, OK 74078, USA\\}

\end{center}

\vspace*{0.1in}
\begin{abstract}

New physics at the TeV scale  is highly anticipated at  the LHC. New particles with color, if within the LHC 
energy reach, will be copiously produced. One such particle is a diquark, having the quantum numbers of 
two quarks, and can be a scalar or a vector. It will decay to two light quarks, or two top quarks, or a top 
and a light quark, (up or down type depending on the quantum number of the produced diquark). If 
singly produced, it can be looked for as a dijet resonance, or as giving extra contribution to the single 
top production or $tt$ production.  
In this work, we consider a color sextet vector diquark having the quantum number of $(ud)$ type, its 
resonance production , and the subsequent decay to $tb$, giving rise to excess contribution to the single 
top production. Even though the diquark mass is large, its strong resonance production dominate the 
weak production of $tb$ for a wide range of the diquark mass. Also its subsequent decay to $tb$ produce 
a very hard $b$-jet compared to the usual electroweak production. In addition,  the missing energy in the 
final state event is much larger from the massive diquark decays. Thus, with suitable cuts, the final state
with $b ,\bar{b }$ and a charged lepton together with large missing energy stands out compared to the 
Standard Model background. We make a detailed study of both the signal and the background. We find 
that such a diquark is accessible at the $7$ TeV LHC upto a mass of about $3.3$ TeV with the luminosity 
1 fb$^{-1}$, while the reach goes up to about $4.3$ TeV with a luminosity of 10 fb$^{-1}$.

\end{abstract}

\vfill

\section{Introduction}
After more than one year of successful running of the Large Hadron Collider (LHC) at CERN,
the data released by the two experiments, ATLAS and CMS have not only improved on some of the limits set by the Tevatron experiments, 
but has already started giving some insights into the TeV scale. So far the results have proved to be consistent 
with predictions from the  Standard Model (SM) with not much deviation which means that the 
LHC data is already pushing the energy frontier of any Beyond Standard Model (BSM) physics predictions. 
As expected the LHC data would  be most sensitive to the strongly interacting sector with 
strong limits obtained from resonant searches of new physics exchanged in the s-channel. As the initial 
states at hadron colliders are colored particles, the most dominant contributions would be through new 
colored resonances as their couplings would be typically of the order of $\alpha_s$.  Such colored particles 
are predicted in many class of BSM theories such as `` squarks" in R-Parity violating supersymmetric 
theories \cite{Barbier:2004ez},  ``diquarks" in super-string inspired $E_6$ grand unification 
models \cite{Hewett:1988xc}, ``excited quarks" in composite 
models \cite{DeRujula:1983ak, Kuhn:1984rj, Baur:1987ga, Baur:1989kv}, 
models with color-sextet fermions \cite{Frampton:1987dn, Martin:1992aq}, 
color-octet vectors such as axigluons \cite{Frampton:1987dn, Bagger:1987fz} 
and colorons \cite{Hill:1991at, Hill:1993hs, ddn:1995pr,Sayre:2011ed}, models with 
color-triplet \cite{Babu:2006wz},
color-sextet \cite{Pati:1974yy, Mohapatra:1980qe, Chacko:1998td} or color-octet 
scalars \cite{Hill:2002ap, Dobrescu:2007yp}.  

These resonant states when produced will decay to two light jets leading to modification of the dijet 
differential cross-section at large invariant mass.  Both ATLAS and CMS collaborations have looked 
at the dijet signal and already put strong constraints on such resonances \cite{Aad:2010bc, 
Khachatryan:2010jd, Aad:2011fq, Chatrchyan:2011ns}.  The most current bounds reported by the 
ATLAS experiment with data corresponding to 1.0 fb$^{-1}$ integrated luminosity at 95\% C.L. 
are $2.99$ TeV for excited quarks, $3.32$ TeV for axigluons and $1.92$ TeV for color-octet
scalars \cite{Aad:2011fq} while the CMS collaboration with the same amount of 
data reports lower bounds of $3.52$ TeV for the 
$E_6$ diquarks ,  $2.49$ TeV  for the excited quarks, $2.47$  TeV for the 
axigluons and colorons.\cite{Chatrchyan:2011ns}. 
 
We note that another interesting prospect other than the dijet signal at LHC would be the modifications 
to the top quark signal due to exchanges from such colored particles. We are interested in particular with
particles of the ``diquark" type which carry non-zero baryon number and couple to a pair of quarks or
anti-quarks. A lot of studies exist in the literature for such diquarks and their resonant effects in 
the dijet signals \cite{Atag:1998xq, Arik:2001bc, Cakir:2005iw, Mohapatra:2007af, Berger:2010fy, 
Han:2010rf, Giudice:2011ak, Richardson:2011df} and pair production of top 
quarks \cite{Barger:2006hm, Frederix:2007gi, Zhang:2010kr, Kosnik:2011jr } at the LHC. 
Scalar triplet diquark contributions have been previously considered in single top quark production 
at the LHC \cite{Gogoladze:2010xd}.
We focus on the case of vector {\it diquarks} which are sextets of $SU(3)_C$ with
charge $Q_e =\frac{1}{3}$. Such particles will be copiously produced
as s-channel resonances  and thus contribute to the dijet final state. It is worth noting that these 
diquarks will also contribute to the single top quark production leading to significant enhancement 
in the production cross-section for the process which is the main thrust of this work.  
An important property of the top, in contrast to lighter quarks, is that it decays before 
hadronization and thus the single top quark production at the LHC can prove to be an ideal 
channel to probe for new physics.  

For our study of the vector diquark we follow the formalism presented in Ref~\cite{Han:2010rf}.
In Sec. \ref{sec:formalism} we present the formalism and give the basic interaction Lagrangian 
relevant for our study and in Sec. \ref{sec:pheno} we discuss the single top production 
cross-section at the LHC and give our results for the signal coming from the diquark 
exchange and present a detail analysis by comparing the signal with the SM background 
for the single top channel. In Sec. \ref{sec:lhc sensitivity} we discuss the LHC reach for diquark 
contribution in the single top channel and in Sec. \ref{sec:concl} we give our conclusions 
with future outlook.

\section{Formalism} \label{sec:formalism}
We are interested in new elementary particles that couple to a pair of quarks directly which would imply
that they carry exotic baryon number. As the LHC is a proton-proton machine, the initial states 
comprised of the the valence quarks would lead to enhanced flux in the parton distributions for
the collision between a pair of valence quarks such as $uu, dd$ or $ud$. Any new particle that couples
to these pairs would carry a baryon number $B=\dfrac{2}{3}$ and will be charged under the SM color 
gauge group $SU(3)_C$. Such states are generally referred to as ''diquarks". We follow the formalism
presented in Ref~\cite{Han:2010rf} where the states are classified according to their charges under the 
SM gauge group $SU(3)_{C}\times SU(2)_{L}\times U(1)_{Y}$ and their spin ($J$). 
We follow the notation of group structure
\begin{equation}
(SU_3,SU_2)_{Q_e}^J,
\label{eq:note}
\end{equation}
where $Q_e$ indicates the electric charge ($T_{3}+Y$). The colored exotics (diquarks) that 
couple to the valence quark pairs can be either color anti-triplets or sextets. Writing them in the 
notation given by Eq. \ref{eq:note}, we have 
\begin{eqnarray}
&& \Phi \sim  ({\rep3}\oplus\bar{\rep6},\rep3)^0_{-4/3,2/3,-1/3},\quad
\Phi_q \sim ({\rep3}\oplus\bar{\rep6},\rep1)^0_q \ \ (q=-1/3,\ 2/3,\ -4/3),
\nonumber \\
&&
A_{U}^\mu \sim ({\rep3}\oplus\bar{\rep6},\rep2)^1_{-1/3,-4/3}\quad
A_{D}^\mu \sim ({\rep3}\oplus\bar{\rep6},\rep2)^1_{2/3,-1/3},
\label{eq:33}
\end{eqnarray}
plus their charge conjugates.
The gauge invariant Lagrangian describing the interaction of the above states is given by \cite{Han:2010rf}
\begin{eqnarray}
\mathcal{L}_{qqD}
& \sim & K^j_{ab}\ \left[ y_{\alpha\beta}\ \overline{Q^C_{\alpha a}}i\sigma_2{\Phi^j}Q_{\beta b} +
\kappa_{\alpha\beta}\ {\Phi^{j}_{-1/3}}\overline{Q^C_{\alpha a}}i\sigma_2Q_{\beta b} \right.   \nonumber\\
&& + \lambda^{1/3}_{\alpha\beta}\ \Phi^{j}_{-1/3}\overline{D^C_{\alpha a}}U_{\beta b}
+\lambda^{2/3}_{\alpha\beta}\ \Phi^{j}_{2/3}\overline{D^{C}_{\alpha a}}D_{\beta b}
+\lambda^{4/3}_{\alpha\beta}\ \Phi^j_{-4/3}\overline{U^C_{\alpha a}}U_{\beta b} \nonumber\\
&&
\left. +\lambda^U_{\alpha\beta}\ \overline{Q^C_{\alpha a}}i\sigma_2\gamma_\mu{A^{j}_U}^\mu U_{\beta b}
+\lambda^D_{\alpha\beta}\ \overline{Q^{C}_{\alpha a}}i\sigma_2\gamma_\mu{A^{j}_D}^\mu D_{\beta b} \right] + \mathrm{h.c.},
\label{diquark.EQ}
\end{eqnarray}
where $\Phi^j = {1\over 2}\sigma_{k} \Phi_{k}^{j}$ with $\sigma_{k}$ the $SU(2)_{L}$ Pauli matrices and
$K^j_{ab}$ are $SU(3)_{C}$ Clebsch-Gordan coefficients with  the quark color indices
$a,b=1-3$, and the diquark color index $j=1-N_D$.   $N_D$ is the dimension of the ($N_D=3$) triplet or ($N_D=6$) antisextet representation.
$C$ denotes charge conjugation, and $\alpha,\beta$ are the fermion generation indices.
The color factor $K^j_{ab}$ is symmetric (antisymmetric) under $ab$ for the ${\rep{6}}\ (\bar{\rep{3}})$ representation.

After electroweak symmetry breaking, the states in Eq.~(\ref{eq:33}) mix and reclassify themselves 
according to color ($\rep3,\ \bar{\rep6}$) and electric charges ($-4/3,\ 2/3,\ -1/3$). 
We are only interested in the couplings of the vector diquark and thus keep only terms 
that involve the vector fields which we denote as $V_{2U}^{N_D},V_U^{N_D},V_D^{N_D}$, where the 
subscripts $2U, U$, and $D$ in the fields indicates 
their electric charge $|Q|$ of two up type quarks, one up and one down type quark respectively.
The relevant interactions among the physical vector states (diquarks) are then described by the 
following effective Lagrangian density 
\begin{eqnarray}
\mathcal{L}_{qqD}^{V} =  K_{ab}^{j} \left[  \lambda_{\alpha \beta}^{'2U} V^{j \mu}_{2U}     
                                      \overline{u^c}_{\alpha a} \gamma_{\mu}P_R u_{\beta b} 
  + \lambda_{\alpha \beta}^{'U} V_{U}^{j\mu} \overline{d^c}_{\alpha a}\gamma_{\mu} P_R d_{\beta b} 
  +  \lambda_{\alpha\beta}^{'D} V_{D}^{j\mu} \overline{u^c}_{\alpha a} 
\gamma_{\mu}P_{\tau}d_{\beta b} \right]+ \mathrm{h.c.} \label{eq:Lagrn}
\end{eqnarray}
%
%
where $P_{\tau}=\frac{1}{2}(1\pm\gamma_{5})$ with $\tau=L,R$ representing left and right chirality projection operators and superscript $ \mu $ is the Lorentz four vector index. Note that we have suppressed
the dimension index ($N_D$) which is common to all the states as the interaction for
the triplet and antisextet are similar. The more general form of the Lagrangian can be found in 
Ref~\cite{Han:2010rf}. 

The coupling $\lambda$'s involved in Eq. \ref{eq:Lagrn} are completely arbitrary and
can be large as long as they 
remain perturbative $\left(\frac{\lambda^{'2}}{4\pi} < 1 \right)$. However most of them are tightly 
constrained by flavor physics. In our case we find that the more stringent constraints come from 
collider data such as the Tevatron and LHC. We focus on the contribution coming from the
interaction vertex that involve the vector diquark $V_D^{\mu}$ which will mediate the 
production of top quark at the LHC and modify the event rates for single top quark production. 
The diquarks can couple to the initial state valence partons in the proton  and are only constrained by 
their coupling strengths. Thus they will contribute to the dijet production at the LHC 
which is enhanced for lighter mass and will be possible to study through the invariant 
mass distribution in the dijet channel. Both the CMS \cite{Chatrchyan:2011ns} and 
ATLAS \cite{Aad:2011fq} experiments at LHC have looked for such narrow resonances 
in the dijet channel and effectively lead to strong constraints on such massive 
resonances. It is worth noting that these colored states do not have direct coupling to 
a pair of gluons and thus the production cross section for such particles is limited by the 
flux of the initial partons in the proton at the LHC. 

\section{Single top production at the LHC} \label{sec:pheno}

\begin{figure}[h]
\centering
\includegraphics[width=34mm]{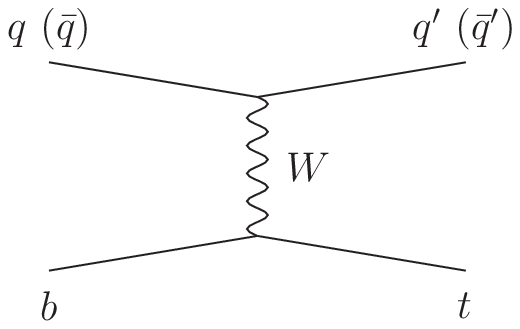}
\hspace{8mm}
\includegraphics[width=34mm]{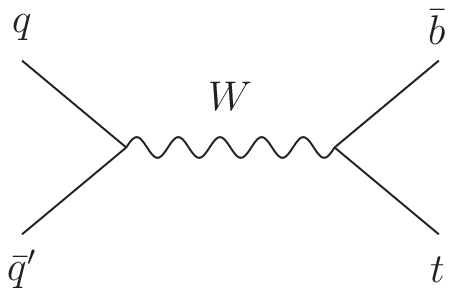}
\hspace{8mm}
\includegraphics[width=70mm]{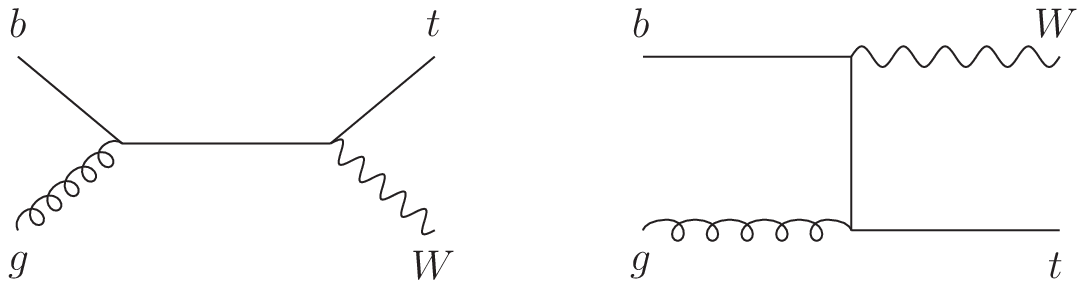}
\caption{The leading order (LO) Feynman diagrams for the  $t$-channel, $s$-channel, and $tW$ production mode for the single top at hadron colliders.}
\label{tsWlo}
\end{figure}

We study the single top-quark production at the LHC where the new physics 
effects on the single top production come from the exchange of the vector 
diquark ($V_D$) in the s-channel. The single top production in the SM can 
proceed in three different channels as shown in Fig.~\ref{tsWlo}.
We can have the (a) $t$-channel production of the top quark with a light quark
in the final state, (b) $s$-channel mode where the top quark is produced with
$\bar{b}$-quark via exchange of a $W$-boson and (c) the associated production 
of the top quark with the $W$-boson via initial state gluon-bottom fusion. 
The $t$-channel processes, $qb \rightarrow q' t$ and 
${\bar q} b \rightarrow {\bar q}' t$, 
are significantly large than
the $s$-channel processes,  $q{\bar q}' \rightarrow {\bar b} t$ at LHC.
The associated production of a top quark with a $W$ boson, 
$bg \rightarrow tW^-$, has a smaller cross-section compared to the 
$t$-channel processes, but is significantly larger than the $s$-channel processes.
The approximate theoretical cross-sections  for the $7$ TeV LHC for the single top processes 
calculated at NNLO \cite{NKtch, NKsch, NKtW} are shown in 
Table~\ref{singletop-csec}.
\begin{table}[!h]
\begin{center}
\begin{tabular}{|c|c|}
\hline {\bf Channel} & NNLO cross-section (pb) \\ 
\hline $t$-channel top & $41.7 {}^{+1.6}_{-0.2} \pm 0.8$ \\ 
\hline $t$-channel anti-top & $22.5 \pm 0.5 {}^{+0.7}_{-0.9}$ \\ 
\hline $s$-channel top & $3.17 \pm 0.06 {}^{+0.13}_{-0.10}$ \\ 
\hline $s$-channel anti-top & $1.42 \pm 0.01 {}^{+0.06}_{-0.07}$ \\
\hline $tW^-$ & $7.8 \pm 0.2 {}^{+0.5}_{-0.6}$\\
\hline
\end{tabular}
\caption{The approximate NNLO cross-sections for single top (anti-top) production of 
mass $m_t=173$ GeV at the LHC with 7 TeV center-of-mass energy. Note that the cross section 
for ${\bar t}W^+$ production is identical to that for top.} \label{singletop-csec}
\end{center}
\end{table}

\begin{figure}[h]
\centering
\includegraphics[width=50mm]{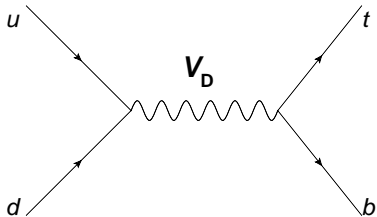}
\caption{ Feynman diagram for the  $s$-channel contribution of the vector 
diquark for the single top production at LHC.}
\label{stVdq}
\end{figure}

For our purposes, we are interested in the corrections to the $s$-channel 
production mode for the single top through the vector diquark exchange.
The Feynman diagram contributing to the production mode is shown in 
Fig.~\ref{stVdq}.
Note that the final state has a $b$-quark instead of the $\bar{b}$ unlike the $s$-channel
mode in SM. The anti-top production through the charge-conjugated vertices would be significantly 
suppressed and different because of the small flux 
of the anti-partons in the colliding protons. Although the top and anti-top
production can be distinguished through the charged lepton identification
coming from the decay, it is practically impossible to distinguish the 
$b$ from the $\bar{b}$ at LHC. Thus we focus only on the production 
mode $pp \to t b$ which will show an effective increase in the $s$-channel
single top production. As the final state distinctly differs from the SM 
subprocess, there are no interference term contributions. Thus we can 
simply evaluate the new physics contributions by calculating the subprocess
$u d \to V^\ast_D \to t b$ at LHC and convolute it with the appropriate 
parton density functions (PDF) for the initial colliding partons.    

The  amplitude for the subprocess  $u d \to V^\ast_D \to t b$ using the
interaction vertices from the Lagrangian as given by Eq.~\ref{eq:Lagrn} is 
\begin{equation}
\overline {|{\mathcal M}|}^2  = \frac{\lambda_{11}^{'2} \lambda_{33}^{'2}}{[(\hat{s}-M_D^2)^2+M_D^2
 \Gamma_{D}^2]}[(\hat{s}-m_{t}^2)(1+\cos \theta)][\hat{s}+m_{t}^2+(\hat{s}-m_{t}^2)\cos \theta]
\end{equation}
where $\theta$ is the scattering angle of the top-quark with the beam axis, 
$M_D$ is the mass of the vector diquark with a decay width of $\Gamma_D$, 
 $\lambda$'s are the Yukawa couplings, $m_t$ is the top quark mass and 
 $\hat{s}=x_1x_2 s$ is the effective center-of-mass energy squared for the
 colliding partons carrying $x_i$ fraction of proton energy. The parton-level Born 
 cross-section then reads
\begin{equation}
\hat{\sigma}(u d\rightarrow t b) = \frac{1}{48 \pi\hat{s}^{2}}\frac{\lambda_{11}^{'2}\lambda_{33}^{'2}}
{[(\hat{s}-M_D^2)^2 + M_D^2\Gamma_{D}^2]}(\hat{s}-m_{t}^2)^{2}(m_{t}^2+2\hat{s})
\end{equation}
\begin{figure}[!t]
\bc
\includegraphics[width=3.5in]{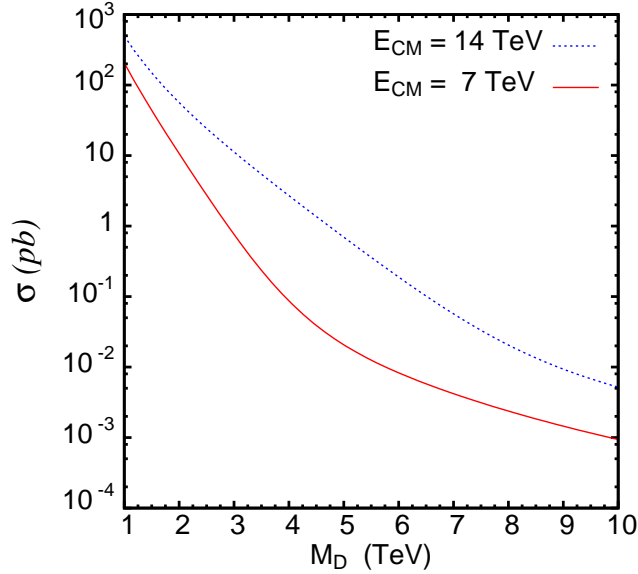}
\caption{\it The production cross-section for $pp \to tb+X$ at LHC as a function
of the diquark mass $M_D$ for two values of the center-of-mass energy,
$E_{CM}=7$ and $14$ TeV. Note that we have chosen $\lambda_{\alpha\alpha}^{\prime}=1$
and $Q=m_t$, the mass of the top-quark.}  \label{csec}
\ec
\end{figure}
The total production cross-section at the LHC for the subprocess through the 
vector diquark exchange in the $s$-channel is given by
\begin{align}
\sigma (pp \to t b +X ) = \sum_{i=1}^{2} \int dx_1 \int dx_2 ~\mathcal{F}_{u_i} (x_1,Q^2) 
\times \mathcal{F}_{d_i} (x_2,Q^2) \times \hat{\sigma} (u_i d_i \rightarrow t b)
\end{align}
where $\mathcal{F}_{u_i}$ and $\mathcal{F}_{d_i}$ denote the PDF's for the up-type
and down-type quarks in the colliding protons, while $Q$ is the factorization scale.
We plot the leading-order production cross section for the process 
$pp \to tb+X$ at LHC for the center-of-mass energies of 7 TeV and 14 TeV
as a function of the diquark mass in Fig. \ref{csec}.  We have set the factorization 
scale $Q=m_t$  and used the {\tt CTEQ6l1} parton density functions 
\cite{Pumplin:2002vw}.  We have used $\lambda_{\alpha\alpha}^{\prime}=1$ 
for evaluating the production cross-section in Fig. \ref{csec}. Note that for 
the diquark mass $M_D < 4$ TeV the contribution is more than $0.1$ pb at the 
current LHC center-of-mass energy for order  $\lambda^\prime =1$ couplings. 
However the couplings are severely constrained for the first two generations 
from the dijet data at the LHC and thus effective cross-sections would be actually 
smaller for the low values of diquark mass.  For the diquarks of mass 
greater than 1 TeV we can assume the top quark as massless and   
in this limit the total decay width  $V_D$ (for decays to $ud + cs+ tu$) is given as
\begin{equation}
\Gamma_D = \frac{\lambda^{'2} M_D}{8 \pi}.
\end{equation}
As the diquark exchange in the $s$-channel would contribute to the dijet 
cross-section there are strong constraints on its mass and coupling strength 
\cite{Chatrchyan:2011ns, Aad:2011fq} . For order one coupling the current 
bound is 3.52 TeV for the $E_6$ diquarks. However, we must 
note that the coupling strength of the diquark to the third generation 
$\lambda_{33}^\prime$ is unconstrained by the dijet data and thus the direct
bounds on the diquark mass can be relaxed for smaller values of the couplings.
The  $\lambda_{33}^\prime$ coupling is however constrained from 
the upper bound on the single-top cross-section in the $s$-channel which is
$< 26.5$ pb at 95\% C.L. using a cut-based analysis \cite{atlas-stop}. For our 
analysis, we have assumed that the couplings to all generations are the same
and thus we use the constraints given by the upper bound on the 
cross-section times branching ratio ($\sigma \times B$) on the dijet 
rate \cite{Chatrchyan:2011ns} to evaluate the cross-sections for different 
values of the diquark mass.  

As pointed out before, we would like to see the effect of the diquark contributions
to the $s$-channel production of single top-quark. To analyze this we
focus on the semileptonic decay mode of the produced top quark leading to the following
final state:
\begin{align}
p p \longrightarrow (t \to b W^+)  b \longrightarrow (W^+ \to \ell^+ \nu_\ell) b b 
\longrightarrow \ell^+ b b \slashed{E}_T
\end{align}
where we restrict ourselves to the choice of $\ell = e, \mu$ for the charged lepton. As
it is very difficult to differentiate between a $b$ and $\bar{b}$ even with the heavy flavor
tagging of the jets in the final state, we are looking at a final state with one positively 
charged lepton ($\ell^+$) and two hard $b$-jets and missing transverse momenta.
A similar final state is expected in the SM from the $s$-channel single top production
as well as various other subprocesses which can lead to similar final state 
topology. At the LHC the significant contributions in the SM to this final state come from the 
following processes:
\begin{align}\label{smbkg}
p~p & \longrightarrow  && W^+Z, ~~t\bar{t},~~ t\bar{b}, ~~W^+ b\bar{b}  \nn \\
      & \hookrightarrow  && tj, ~~W^+jj, ~~(tW^- + \bar{t}W^+), ~~W^+W^-
\end{align}
where the processes in the upper row of Eq. \ref{smbkg} give two $b$-jets while 
the processes in the lower row lead to light quarks in the final state which are 
identified as $b$-jets because of mistag.  For our parton-level analysis, we choose 
a $b$-tagging efficiency of 50\% while the mistag rate for light quarks tagged as 
$b$-jets is taken as 1\%. We must point out that the $b$-tag efficiency and the 
mistag rates are dependent on the transverse momenta ($p_T$) and 
rapidity ($\eta$). Our choice does not include these effects. However, to do 
such detailed analysis one would also need to include various other systematics 
including showering and hadronization effects at the LHC and detector-level 
simulations which is beyond the scope of this work. So we assume that our choice 
for the efficiencies and the mistag rate is a good approximation when averaged over the entire range of 
transverse momenta for the quarks within the allowed rapidity gap.  

To select the particles in the final state we demand that they satisfy some basic  
kinematic selection cuts. 
\begin{itemize}
\item For the two $b$-jets we demand that they have a minimum transverse 
momenta given by $p_T^{b} > 20$ GeV and are within the rapidity gap
$ |\eta_{b}| < 2.5$.
\item The charged lepton ($\ell^+= e^+,\mu^+$) is required to have a minimum 
transverse momenta given by $p_T^{\ell^+} > 20$ GeV and is within the rapidity gap
$ |\eta_{\ell^+}| < 2.5$.
\item The final states must account for a minimum missing transverse momenta 
given by $\slashed{E}_T > 50$ GeV.
\item To resolve the final states in the detector they should be well separated. To
achieve this we require that they satisfy $\Delta R_{ij} > 0.2$ with $i,j$ representing 
the $b$-jets and the charged lepton. The variable $\Delta R_{ij}$ defines the 
separation of two particles in the ($\eta,\phi$) plane of the detector with  
$\Delta R_{ij}=\sqrt{(\eta_i-\eta_j)^2+(\phi_i-\phi_j)^2}$, where 
$\eta$ and $\phi$ represent the pseudo-rapidity and azimuthal angle 
of the particles respectively.  
\item To suppress large contributions of gluon splitting into two ($b$) jets we 
demand that the minimum invariant mass of two $b$-jets satisfy 
$M^{inv}_{b_1b_2} > 10$  GeV. 
\item We also demand that there are no additional jets with $p_T > 20$ GeV or 
additional charged leptons with $p_T>10$ GeV.
\end{itemize}

\begin{figure}[!h]
\includegraphics[width=2.1in]{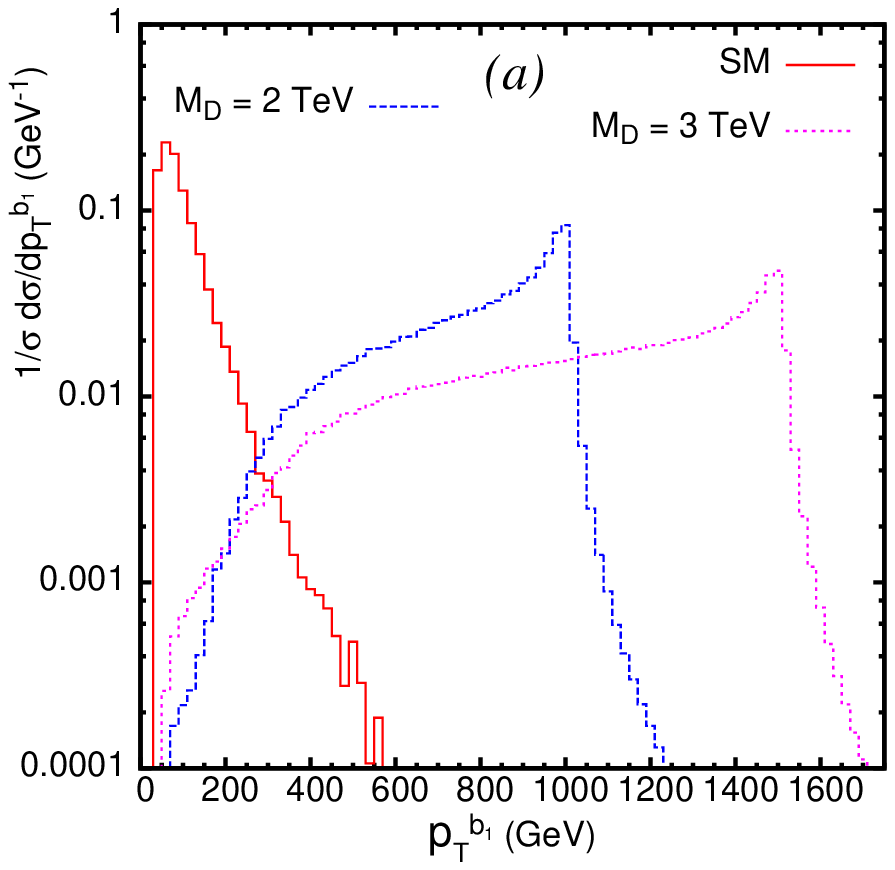}
\includegraphics[width=2.1in]{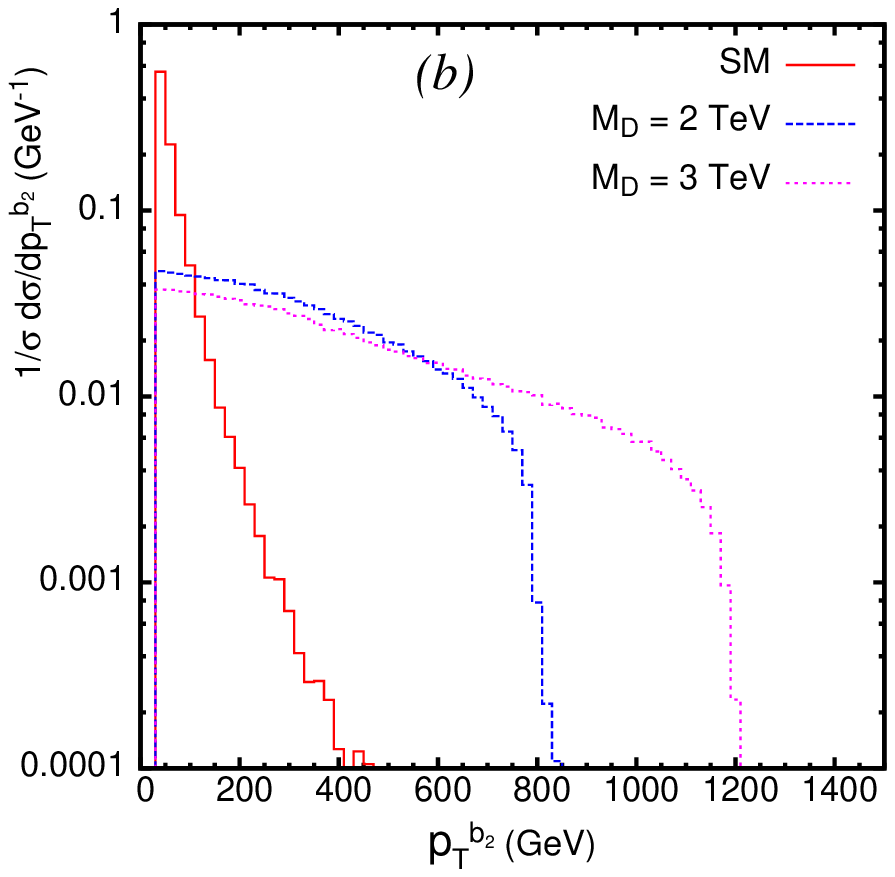}
\includegraphics[width=2.1in]{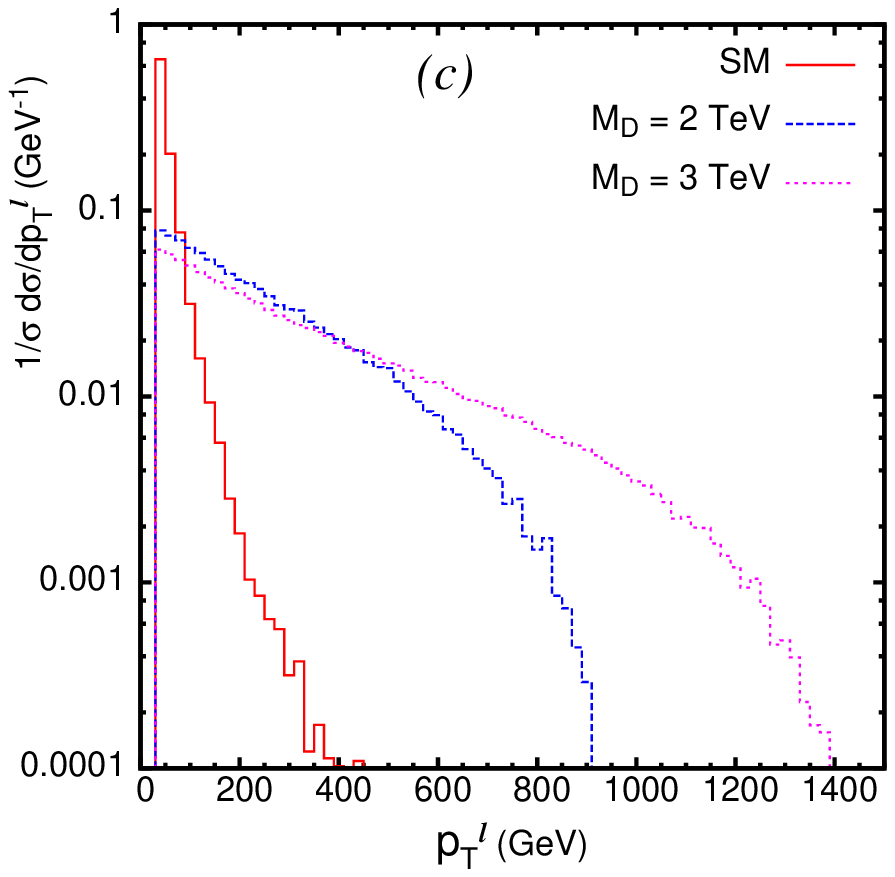}
\caption{\it Normalized differential distribution of the cross section as a function of 
the transverse momenta ($p_T$) of the visible particles in the final state for the SM 
background and the signal for two different values of the mass of the 
diquark $M_{D}=2$ TeV and  $M_{D}=3$ TeV.} \label{fig:pT}
\end{figure}

\begin{figure} [!ht]
\includegraphics[width=2.1in]{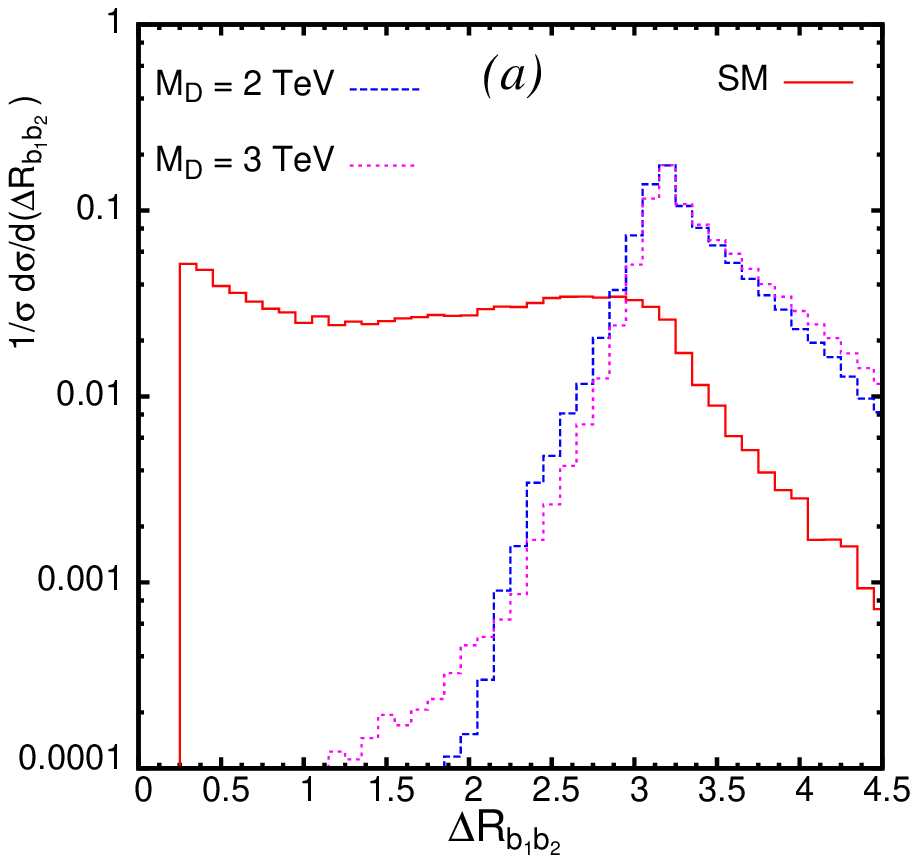}
\includegraphics[width=2.1in]{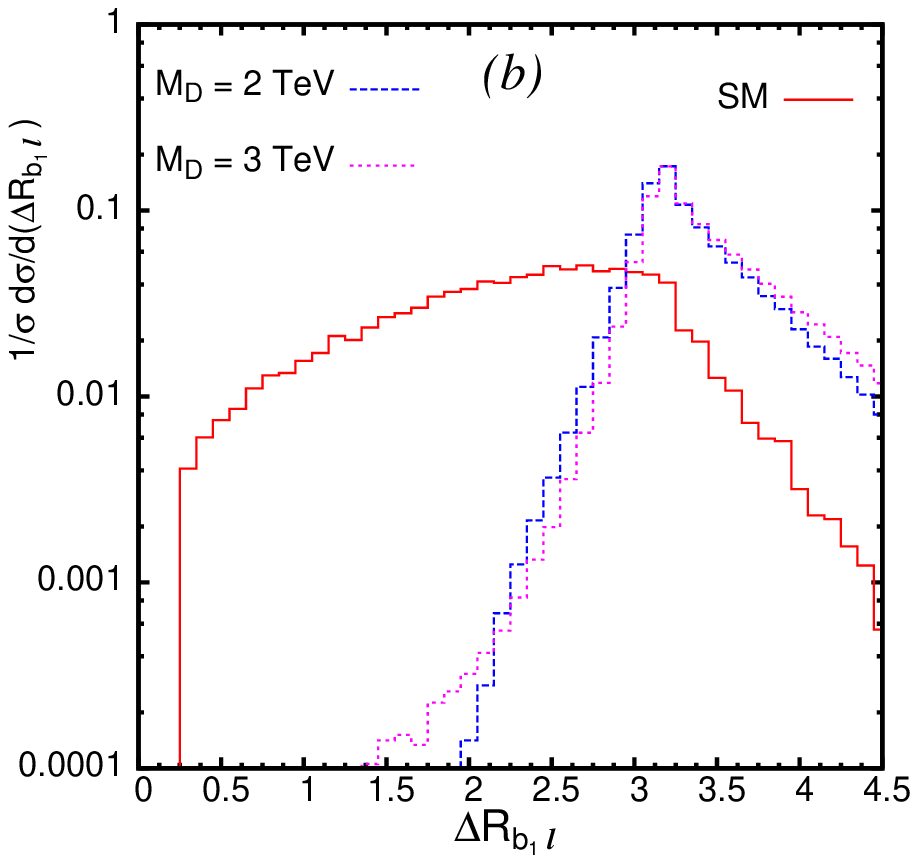}
\includegraphics[width=2.1in]{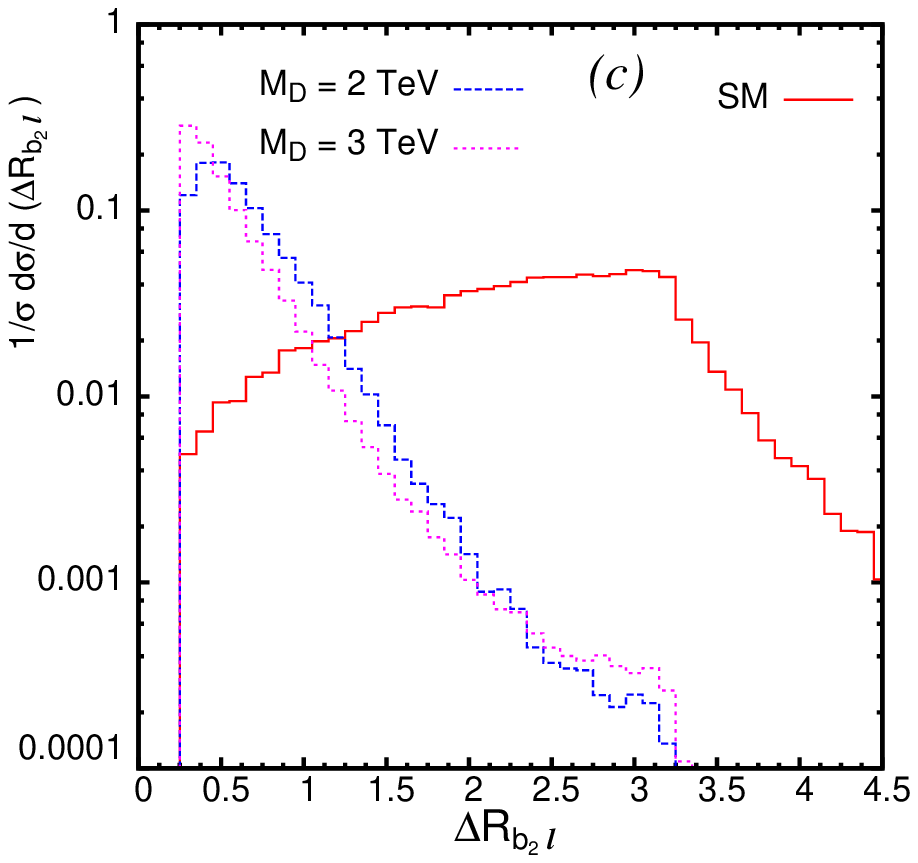}
\caption{\it Normalized differential distribution of the cross section as a function of 
the variable $\Delta R$ illustrating the isolation between two particles in the 
$(\eta,\phi)$ plane of the detector for the SM background and the signal for two 
different values of the mass of the diquark $M_{D}=2$ TeV and  
$M_{D}=3$ TeV.} \label{fig:DR}
\end{figure}

\begin{figure}[!h]
\includegraphics[width=2.1in]{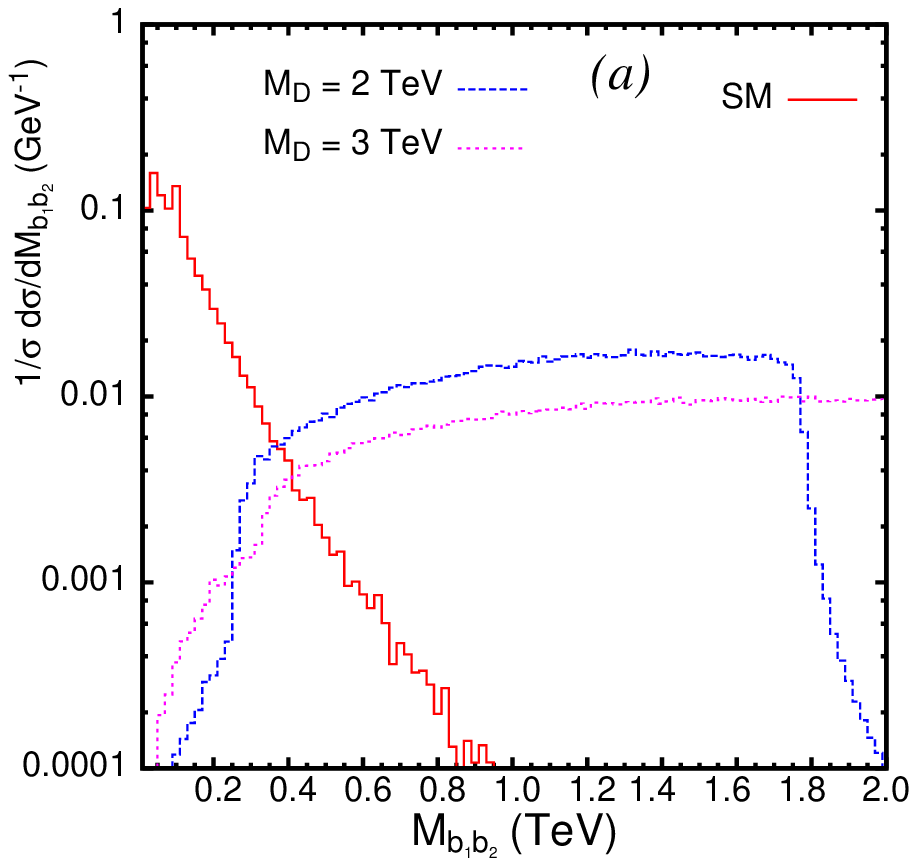}
\includegraphics[width=2.1in]{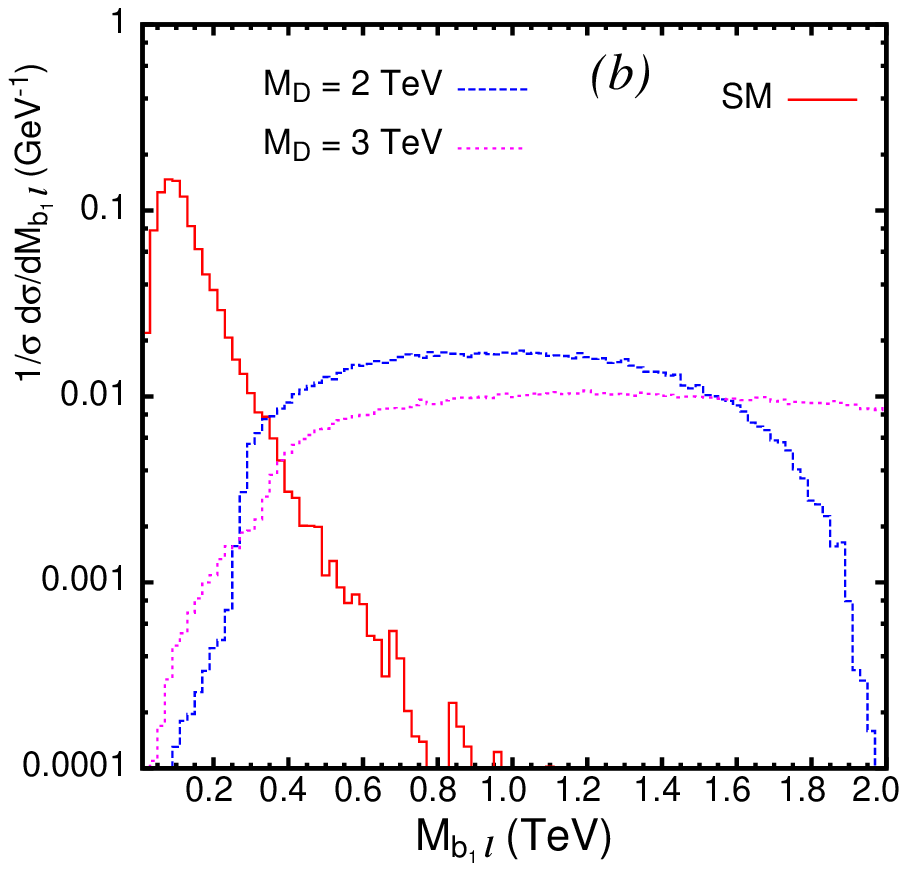}
\includegraphics[width=2.1in]{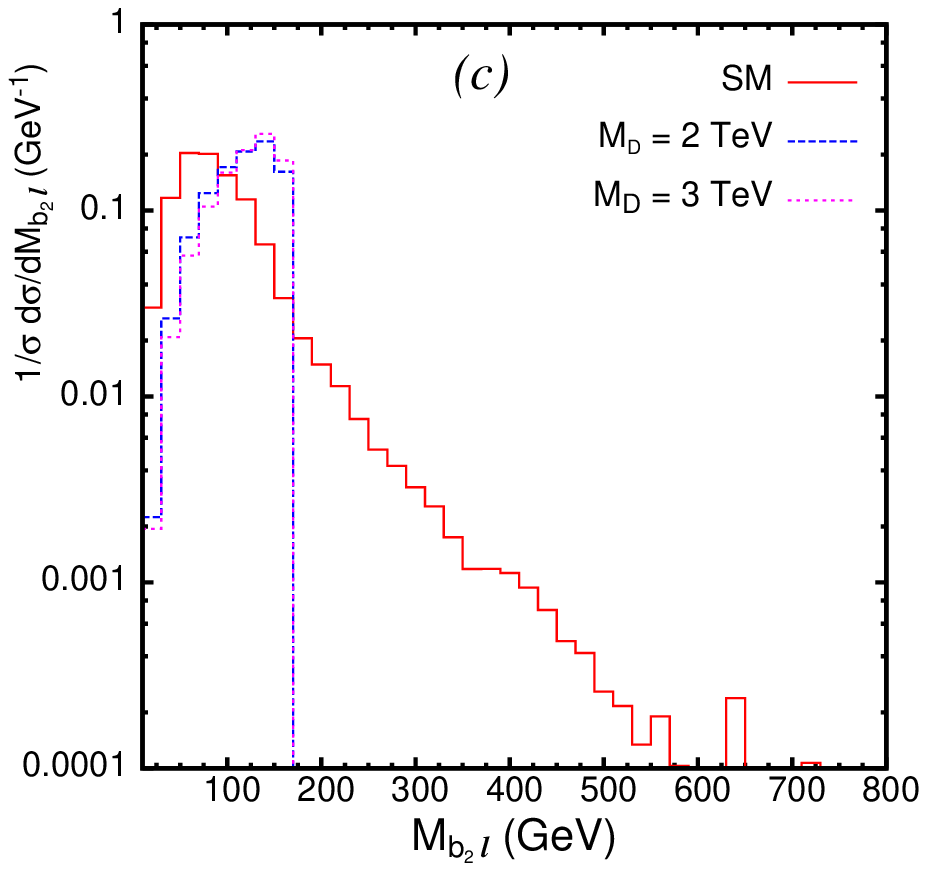}
\caption{\it Normalized differential distribution of the cross section as a function 
of the invariant mass of  a pair of visible particles in the final state for the SM 
background and the signal for two different values of the mass of the 
diquark $M_{D}=2$ TeV and  $M_{D}=3$ TeV.} \label{fig:Minv}
\end{figure}

With these basic selection cuts and efficiency rates for $b$-tagging we evaluate 
the contributions to the final state within the SM using the event generator 
{\sf MadGraph 5} \cite{Alwall:2011uj}. The contributions from the different 
processes are collected in Table \ref{bkgcsec}. We note that while the 
$s$-channel single top quark process gives a significant rate, the most 
dominant contribution comes from the QCD induced process $pp\to W^+ b\bar{b}$.
Summing over all significant contributions we find that the LO SM cross-section 
at LHC with $\sqrt{s}=7$ TeV, for the final state of $\ell^+ + 2 b + \slashed{E}_T$, 
is $\simeq 141$ fb.
\begin{table}[!h]
\begin{center}
\begin{tabular}{|c|c|}
\hline Sub-process & $\sigma(\ell^+ + 2 b + \slashed{E}_T)$  (fb) \\ 
\hline $pp\to WZ$ & 7.33 \\ 
\hline $pp\to Wjj$ & 9.97 \\ 
\hline $pp\to Wb\bar{b}$ & 87.55\\ 
\hline $pp\to tW$ & 0.53 \\
\hline $pp\to t\bar{t}$ & 5.51\\
\hline $pp\to tj$ & 6.24 \\
\hline $pp\to t\bar{b}$ & 23.32\\
\hline
\end{tabular}
\caption{The total SM background cross section for the dominant sub-processes 
contributing to the final state $\ell^+ + 2 b + \slashed{E}_T$ estimated at the 
parton-level, using the event generator {\sf MadGraph 5}.  Note that we have 
included a b-tagging efficiency of 50\% for the $b$-jets and a mistag rate of 1\% 
for the light quarks to be tagged as $b$-jets for the final cross sections shown in 
the table.} \label{bkgcsec}
\end{center}
\end{table}
We calculate the contributions for the above final state coming from the single
top production via the exchange of a vector diquark in the $s$-channel for two 
different values of the diquark mass, {\sl viz.} $M_D=2$ TeV and $M_D=3$ TeV.
The effective couplings (assuming equal couplings to all generations) for the two 
choices of the diquark mass are restricted by the dijet data and given by the
upper bound on $\sigma \times B (pp \to qq) < 0.121$ pb for $M_D=2$ TeV 
and $\sigma \times B (pp \to qq) < 0.021$ pb for $M_D=3$ TeV which lead to
the upper bound on the effective $\lambda^{\prime}$ coupling as $0.24$ and
$0.3$ respectively. Using the maximum allowed couplings, we estimate the 
cross-section $\sigma(\ell^+ + 2 b + \slashed{E}_T)$ with the selection 
cuts and efficiency rates as  $2.95$ fb for $M_D=2$ TeV and $0.42$ fb for 
$M_D=3$ TeV. However, we must note that these upper bounds are relaxed
as there are no severe constraints on the diquark coupling to the third generation 
quarks. The dominant constraint comes from the upper bound of the $s$-channel single 
top-quark production cross-section at LHC \cite{atlas-stop} which puts an upper bound on 
the product of the couplings as 
$\lambda_{ii}^{\prime} \lambda_{33}^{\prime} \leq 0.45$. This allows the 
$\lambda_{33}^{\prime} \simeq 1$ which would imply that the 2 TeV and 3 TeV 
diquark contributions to the $\ell^+ + 2 b + \slashed{E}_T$ final state can be as 
large as  $51$ fb and $4.7$ fb respectively. 
The relative significance of the diquark contributions can be enhanced by 
selecting events through additional kinematic cuts suited to suppress the SM 
background without effecting the new physics signal. To understand the 
kinematic characteristics of the final state events coming from the SM processes and
the diquark mediated process we plot the normalized differential distributions of the 
cross-section as a function of a few kinematical variables in Fig. \ref{fig:pT}-\ref{fig:Minv}.  

In Fig. \ref{fig:pT} we plot the normalized differential distribution of the cross-section with 
respect to the transverse 
momenta ($p_T$) of the visible particles in the final state, {\sl viz.} the two $b$-jets and the charged 
lepton for the combined SM background and for the individual diquark contributions of mass 
$M_D=2$ TeV and $3$ TeV respectively. The $b$-jets are arranged according to their $p_T$'s and 
so the leading $b$-jet ($b_1$) from the diquark mediated process comes almost always from the 
decay of the heavy  diquark and thus peaks in the high $p_T$ region when compared to the SM 
distribution which falls very rapidly as the $b$-jets get harder. As the diquark mass is increased 
we can see that the $b$-jet $p_T$ peaks at much higher value (corresponding to the Jacobian 
peak at $\simeq \frac{M_D}{2}$). Again for the diquark mediated process the sub-leading 
$b$-jet ($b_2$)  comes from the decay 
of a  highly boosted top-quark and thus carries large $p_T$ as shown in Fig. \ref{fig:pT} (b). 
The charged lepton ($\ell^+$) also shows similar behavior like the sub-leading $b$-jet as 
it also carries the boost of the top-quark from which it originates. 

In Fig. \ref{fig:DR} we plot the normalized differential distribution of the cross-section with respect to the 
isolation parameter $\Delta R$ between a pair of visible particles. Note that the characteristic 
behavior of particles coming from the decay of heavy resonance would be that they are produced 
back-to-back in the transverse plane and are well separated which would imply peaking at large 
values of $\Delta R$. This is the feature expected between the leading $b$-jet  ($b_1$) and 
sub-leading  $b$-jet  ($b_2$) as well as the charged lepton ($\ell^+$) when the 
events correspond to the process mediated by the diquark as evident in Fig. \ref{fig:DR} (a) and (b). 
However, when the pair of particles originate from the decay of the highly-boosted top 
quark ($b_2,\ell^+$) the relative separation between them would be much smaller as seen in 
the distributions shown in Fig. \ref{fig:DR}(c). The SM background does not have these characteristic
features and shows a relatively uniform distribution in the $\Delta R$ variable because of contributions
from multiple processes smoothening away any distinct effects of an individual contribution.

Another important kinematic variable relevant for any resonant production mode is the 
invariant mass of a pair of visible particles which reconstructs the mass of the resonant particle. In
our case the heavy diquark decays to $t b$ where the top decays further semileptonically.  Thus
we focus on the invariant mass distributions of the visible decay products. In Fig. \ref{fig:Minv}
we plot the normalized differential distribution of the cross-section with respect to the invariant mass between a 
pair of visible particles. We  note that the reconstructed invariant mass 
$M_{bb\ell\slashed{E}_T}$ will give us a good but approximate estimate of the diquark mass, which we 
have not shown here. As shown in Fig. \ref{fig:Minv} (a) and (b), the distributions for the diquark
mediated process for the invariant mass of the two $b$-jets $(b_1b_2)$  and the 
leading $b$-jet with the charged lepton $(b_1\ell)$ would peak for larger values and 
drop rapidly as it approaches the diquark mass which acts
as an upper bound for the invariant mass value. The SM contributions however fall off very rapidly
for large invariant mass in these particle pairs which is remnant of the fact that there are no 
heavy TeV mass particle in SM that can cause such a behavior. However, the invariant mass distribution
for the sub-leading $b$-jet and the charged lepton has an upper cut-off at the top mass ($m_t$) for
the diquark mediated process since the two particles almost always originate from the decay of the
top quark and so the invariant mass of the decay products can never be greater than the mass of the
mother particle. This can be seen in Fig. \ref{fig:Minv} (c). Note that the SM distribution has a long 
tail as there are other contributions which get superposed on the single top channel contribution 
making it significantly different from that of the diquark mode.   

The dependence of the differential cross section on the different kinematic variables gives us a 
hint as to what should be the additional kinematic selection on the final state events which 
would make the diquark effects stand out against the SM contributions. We find that the most effective
selection turns out to be the transverse momenta of the leading $b$-jet which originates from 
the decay of the diquark itself. We find that a more stringent requirement that the leading $b$-jet
satisfies $p_T > 400$ GeV is very helpful in achieving a greater suppression of the SM background
without affecting the diquark contributions significantly.  With this additional cut, the SM background
is reduced to $\simeq 0.59$ fb while the contribution for the diquark with mass $M_D=2$ TeV is 
reduced only by 7\%  and  that with mass $M_D=3$ TeV is reduced only by 4\%, thus significantly 
increasing the sensitivity to the diquark effects in the single top channel. The other promising 
variables are the $\Delta R$ and the invariant mass distribution of the two $b$-jets and that 
between the leading $b$-jet and the charged lepton which show significant difference between 
the diquark contribution and the SM background and will increase the signal significance with appropriate selection cuts when combined with the $p_T$ selection. 

\section{LHC Sensitivity}\label{sec:lhc sensitivity}

\begin{figure}[!h]
\includegraphics[width=2.8in]{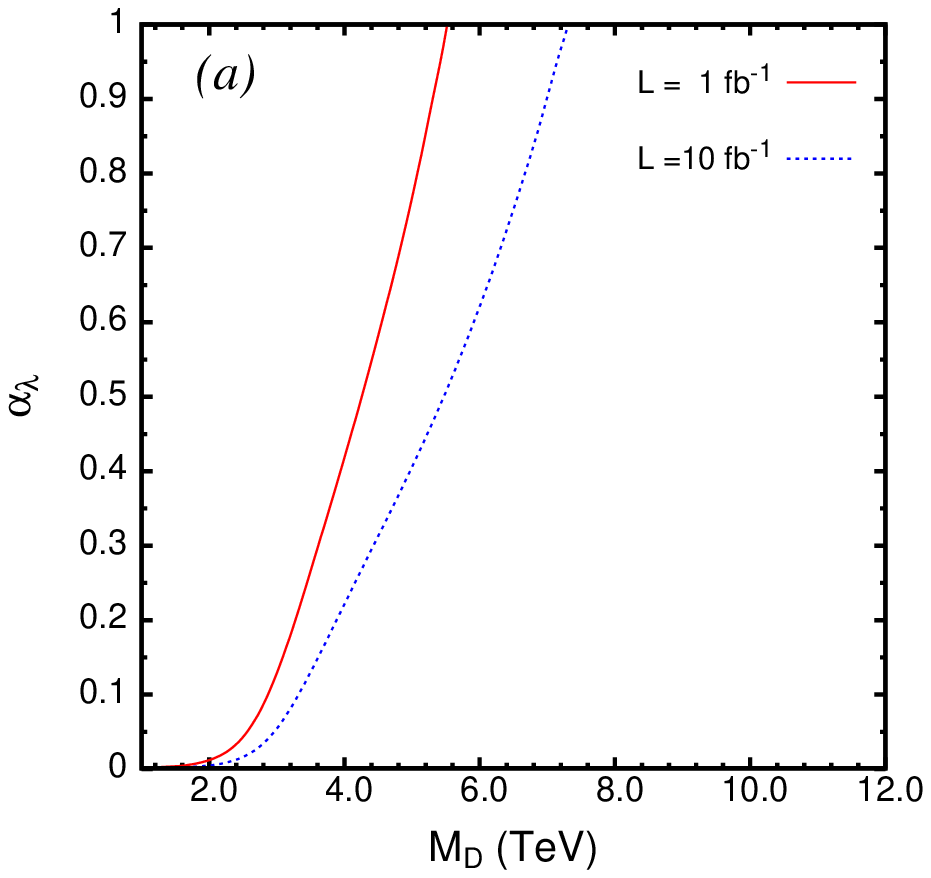}
\includegraphics[width=2.8in]{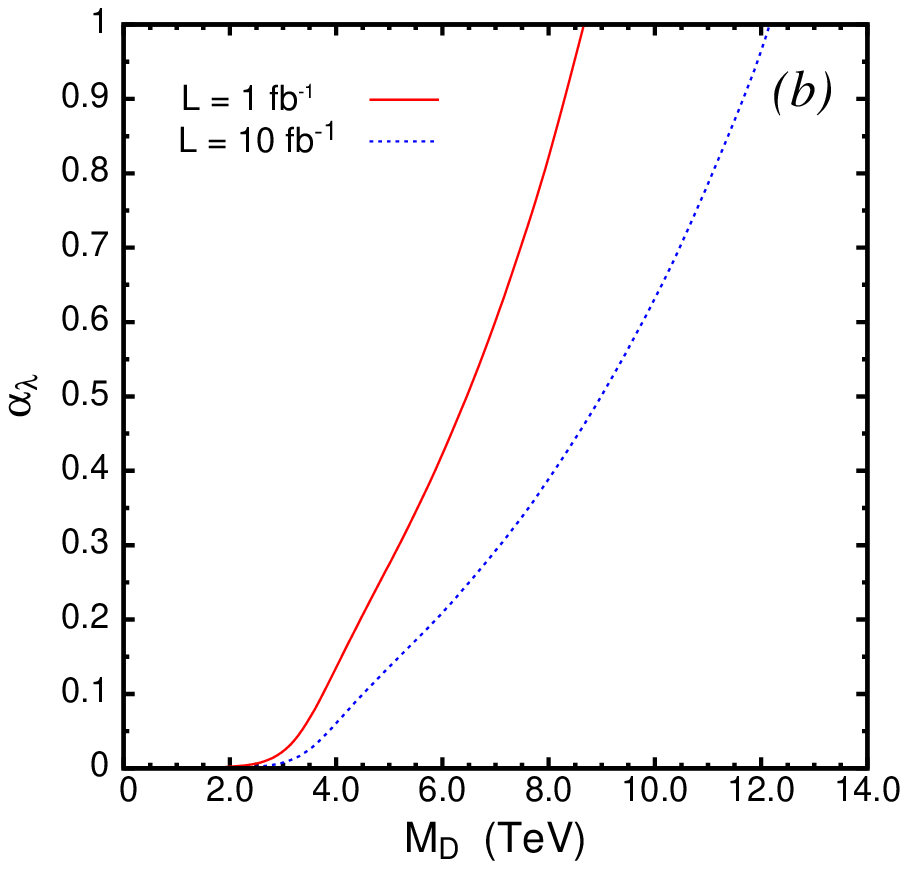}
\caption{\it Illustrating the reach for the LHC running at 7 TeV center-of-mass
energy in the single top channel for the diquark mass and its coupling to SM
quarks. The reach is shown for two values of the integrated luminosity where
(a) shows the reach with the basic cuts used for the single top analysis while
(b) shows the reach with a stronger cut of $p_T > 400$ GeV on the leading b-jet.} \label{significance}
\end{figure}

We  have simulated the events for diquark contribution and the SM background for the final 
state $\ell^+ + 2 b + \slashed{E}_T$ at LHC with $\sqrt{s}=7$ TeV for two values of the 
integrated luminosity of 1 fb$^{-1}$ and 10 fb$^{-1}$. The statistical significance is calculated 
using \cite{statsig}
\begin{equation}
\mathcal{S} = \sqrt{2\times \left[(s+b) \mathrm{ln}(1+\frac{s}{b}) -s\right]}
\end{equation}
where $s (b)$ is the number of diquark (SM background) events for the corresponding 
integrated luminosity. We show the sensitivity of the LHC  to the diquark mass as a function 
of its coupling in Fig. \ref{significance}. We have assumed that the vector diquark couples to all 
generations of SM quarks with the same strength such 
that $\alpha_\lambda \equiv \lambda^{'2}_{ii} /4\pi$. 
However $\alpha_\lambda$ in Fig. \ref{significance} can  also  be redefined as 
$\alpha_\lambda \equiv \lambda^{'}_{ii} \lambda^{'}_{33}/4\pi$  where 
$ \lambda^{'}_{ii}= \lambda^{'}_{11}= \lambda^{'}_{22}$ which means that we can treat 
$\lambda^{'}_{33}$ which is unconstrained by dijet data separately. In Fig. \ref{significance} (a)
we show the sensitivity of LHC in the $(\alpha_\lambda , M_D)$ parameter plane 
at $3\sigma$ statistical significance using the basic selection cuts discussed in 
Sec. \ref{sec:pheno}. For $\lambda^{'2}= 4\pi\alpha_\lambda \sim \mathcal{O}(1)$, we find that the
LHC can rule out diquark of mass $M_D\simeq 2.8$ TeV with data corresponding to 1 fb$^{-1}$
of integrated luminosity while the reach becomes $M_D\simeq 3.65$ TeV with data 
corresponding to 10 $fb^{-1}$ with the same set of cuts. This reach is improved if we include
the strong cut on the transverse momenta of the leading $b$-jet of $p_T>400$ GeV, which 
we show in Fig. \ref{significance} (b). We find that with an integrated luminosity of 1 fb$^{-1}$
which is already collected by each of the two experiments at LHC, the vector diquark mass accessible 
to LHC at  $3\sigma$ statistical significance would be $M_D \simeq 3.25$ TeV while the 
reach expands to $M_D \simeq 4.3$ TeV with $L=10$ fb$^{-1}$. Thus we find that the single top
channel with the appropriate cuts competes quite well with the dijet channel and can prove to
be an improvement over the dijet channel if the coupling of the diquarks to the third generation
quarks is larger than its coupling to the first two generations.   

\section{Summary and Conclusions} \label{sec:concl}
In this work we have studied the single top quark production at the LHC running at 
$\sqrt{s}=7$ TeV. We have considered the effects of a vector diquark resonance on the 
single top quark production ($pp \to tb$) and compared it with the SM background 
for the $\ell^+ 2b ~\slashed{E}_T$ final state, where the top decays semileptonically.
We find that there could be large enhancements in the single top quark production 
at LHC and find that various kinematic distributions effectively capture the essence of 
a heavy particle exchange which distinguish it from SM background. We have shown that the
use of specific selection cuts on the kinematics of the final states can lead to improved 
significance for the effects of the diquark exchange and increase the LHC sensitivity
to heavier diquark mass. We also note that for the vector diquark coupling more strongly 
to the third generation quarks than the first two generations, the LHC can be sensitive to 
higher values of the diquark mass when compared to the dijet channel. 

In this work, we have restricted ourselves to the study of the 
sextet vector diquark exchange  and we note that similar results can be also
obtained for the triplet vector diquark or the scalar diquarks \cite{Gogoladze:2010xd} which 
contribute to the single top quark production at the LHC. The vector diquark effects on the
top quark polarization would be different from that of a scalar diquark 
exchange \cite{Zhang:2010kr} as the Lorentz structure at the 
interaction vertices will be different which can prove to be a good discriminant in 
distinguishing the vector from the scalar exchange \cite{Arai:2010ci, Huitu:2010ad}.

\section*{Acknowledgments}

This work is supported in part by the United States Department of Energy, Grant
Numbers DE-FG02-04ER41306 and DE-FG02-04ER46140.

\end{document}